\documentstyle[11pt,aaspp4]{article}

\lefthead{XU, FANG \& WU}
\righthead{Scale Invariance of Rich Cluster Abundance}

\begin{document}

\title{Scale Invariance of Rich Cluster Abundance: 
A Possible Test for Models of Structure Formation}

\author{Wen Xu
 }

\affil{Department of Physics, University of Arizona, Tucson, AZ 85721;
 and Beijing Astronomical Observatory, Beijing 100080, China}

\author{Li-Zhi Fang 
}

\affil{Department of Physics, University of Arizona, Tucson, AZ 85721}

\and

\author{Xiang-Ping Wu\footnote{E-mail address: xu, fanglz, 
wu@time.physics.arizona.edu}}

\affil{Beijing Astronomical Observatory, Chinese Academy of Sciences,
Beijing 100080, China; and Institute of Astronomy, National Central
University, Chung-li, Taiwan 32054, ROC}

\begin{abstract}
We investigate the dependence of cluster abundance $n(>M,r_{cl})$, i.e.,
the number density of clusters with mass larger than $M$ within
radius $r_{cl}$, on scale parameter $r_{cl}$. Using numerical simulations
of clusters in the CDM cosmogonic theories, we notice that the abundance 
of rich clusters shows a simple scale invariance such that 
$n[>(r_{cl}/r_0)^{\alpha}M, r_{cl}]= n(>M,r_0)$,
in which the scaling index $\alpha$ remains constant in a scale range 
where halo clustering is fully developed. The abundances of scale $r_{cl}$ 
clusters identified from IRAS are found  basically to follow this scaling, 
and yield  $\alpha \sim 0.5$ in the range $1.5 < r_{cl} < 4 h^{-1}$Mpc.
The scaling gains further supports from independent measurements of 
the index $\alpha$ using samples of X-ray and gravitational lensing mass 
estimates. We find that all the results agree within error limit as:
$\alpha \sim 0.5 - 0.7$ in the range of $1.5 < r_{cl} < 4 h^{-1}$Mpc.
These numbers are in good consistency with the predictions of OCDM 
($\Omega_M=0.3$) and LCDM ($\Omega_M+\Omega_{\Lambda} =1$), while the 
standard CDM model has different behavior. The current result seems 
to favor models with a low mass density.

\end{abstract}

\keywords{cosmology: theory - galaxies: clusters: general -
  large-scale structure of universe }

\section{Introduction}

It is generally believed that the gravitational clustering of scale free 
initial perturbations is self-similar (e.g. Peebles 1980). The
statistical descriptions of large scale structures, such as correlation 
functions and number densities, show somewhat of a scaling behavior.
Among various observed objects, clusters probably are the best for 
studying this gravitational scaling, because their distribution and 
clustering are dominated by gravitation and less ``contaminated" by hydro 
processes. Therefore, observed clusters can directly be identified as 
massive halos of N-body simulation samples. A multi-scale identification 
of clusters will be useful to reveal the expected scaling. 

Traditionally, clusters are identified from either observed or simulated 
distribution of galaxies or masses within a sphere of given radius $r_{cl}$,
say, the frequently adopted Abell radius $r_{cl}=1.5 h^{-1}$ Mpc where
$h = H_0/100$ km s$^{-1}$ Mpc$^{-1}$. However, {\it a priori} choice of 
$r_{cl}$ is somewhat arbitrary, or dependent on observational limits. 
Clusters have already been identified and studied on very different radius 
$r_{cl}$. For example, a wide range of radius covering from
0.01 - 4.3 $h^{-1}$ Mpc has been used in the current determination of 
cluster masses (e.g. Wu \& Fang 1996, 1997a and references therein).
Therefore, a multi-scale identification of clusters is a necessary 
extension of the ``standard" Abell radius $r_{cl}=1.5 h^{-1}$ Mpc. It 
will provide information, such as the scale invariance of the mass functions 
of clusters, which can not be seen by one scale identification.  
This is the goal of this paper.

As we know, cluster study has been significant for understanding the nature 
and evolution of cosmic structures on scales of $\sim$ 1 - 10 Mpc. In 
particular, cosmological parameters can be well constrained by the observed 
properties of clusters, such as the abundances and correlations (Bahcall 
\& Cen 1992 (BC); Jing et al. 1993; Jing \& Fang 1994; Viana \& Liddle 1996;
Carlberg et al. 1997a; Bahcall, Fan \& Cen 1997), the substructures (Jing 
{\it et al.} 1995), the luminosity-temperature relations (Oukbir \& 
Blanchard 1997) and the gravitational lensing phenomena (Wu \& Fang 1996; 
Wu et al. 1997a). We believe that the multi-scale identification of clusters
and the scaling invariance would be able to add more constraints on relevant
parameters.  

In \S 2, we summarize the theoretical background of the possible scaling 
of the multi-scale identified clusters. In \S 3, we present the 
result of identifying clusters on different scales in three popular structure 
formation models: 1) the standard cold dark matter model (SCDM), 2) low density, 
flat CDM model with a non-zero $\lambda$ (LCDM), and 3) open CDM model (OCDM). 
Using these samples, we show that the mass functions of rich clusters are 
scaling invariant. In \S 4, the predictions of scaling are compared with 
observations of clusters identified from IRAS sample. In \S 5, 
the index of the scaling is detected from samples of X-ray and gravitational 
lensing clusters with mass estimates. Finally, we describe our main 
findings in \S 6.

\section{Theoretical background}

It is generally believed that cosmic gravitational clustering after the 
decoupling can roughly be described by three r\'egimes: 1. linear regime;
2. quasilinear regime which is dominated by scale-invariant radial infall;
and 3. fully developed nonlinear regime dominated by virialized nonradial
motion. Since the amplitudes of the cosmic temperature fluctuations revealed
by COBE are as small as
$\Delta T/T\sim 10^{-6}$ (Bennett et al. 1996), the gravitational clustering
should remain in the linear regime, or at most in the quasilinear regime,
on comoving  scales larger than about 10 $h^{-1}$ Mpc and at redshifts
higher than 2. On the other hand, observations indicate that clusters of
galaxies with size of about 1.5 h$^{-1}$ Mpc are probably the largest
fully developed or virialized objects. Thus, the quasilinear evolution
should be substantial on the scales from about 2 to 10 h$^{-1}$ Mpc.
Obviously, these facts are useful for discriminating among models of
structure formation. Because different regime generally has different 
behavior of scaling, an effective way of doing model discrimination would 
be to compare the model predicted scaling with observations.

In linear regime, the scaling is straightforward. If the primeval density
perturbations are Gaussian and have power law spectrum
$P(k) = |\delta (k) |^2 = Ak^{n}$, the variance of the initial mass
fluctuation within a spherical region on length scale $R$ is given by
$\langle M \rangle_R \propto R^{(3-n)/2}$. Therefore, there is a scaling of
$R \rightarrow R'$, and $\langle M \rangle_{R'} \rightarrow
(R'/R)^{(3-n)/2} \langle M \rangle_{R}$. Since $\langle M \rangle_R$ is
a statistical average with respect to $N(M, R)dM$, 
the number density of
radius $R$ halo with mass from $M$ to $M+dM$, the scaling of
$\langle M \rangle_R$ indicates the scaling of the number density as
\begin{equation}
n[>(R'/R)^{-(3-n)/2}M, R'] = n(>M, R)
\end{equation}
where
\begin{equation}
n(>M,R) \equiv \int_{M}^{\infty} N(M,R)dM
\end{equation}
which is the accumulative number density of radius R halo
with mass larger than M.

Since $R$ is the initial radius of the halos, not the radius $r_{cl}$ of
the developed halos, the scaling Eq.(1) cannot be directly tested by
observed halos. Only in the stage of linear regime, the scale of halos
$r_{cl} \propto R$, we have then scaling relationship as
\begin{equation}
n[>(r'_{cl}/r_{cl})^{-\alpha}M, r'_{cl}] = n(>M, r_{cl})
\end{equation}
where scaling index $\alpha_{L}=(3-n)/2$, subscript $L$ is for linear
regime. Namely, the scaling index $\alpha$ is a function of $n$, or
it is model-dependent.

Beyond linear regime, the existence of scaling behavior of gravitational
clustering can be shown by comparing the gravitational clustering of cosmic 
matter with the evolution of the profile of a growing interface, or
surface roughing. The later is described by the so-called KPZ equation
(Kardar, Parisi \& Zhang 1986), which mainly consists of terms of
nonlinear evolution and stochastic force, or stochastic initial
perturbations. A major breakthrough in this approach has been to
show the existence of universal scaling properties of the surface 
growth by the dynamical renormalization group (Barab\'asi \& Stanley 1995)).
It has been found that the equations describing the evolution of cosmic
density perturbations are essentially KPZ-like (Berera \& Fang 1994).
The equations of cosmic gravitational clustering contains the similar
nonlinear term and a stochastic initial perturbations. The gravitational
potential of cosmic matter plays the similar role as the height $h(x)$ of 
the surface.
Thus, the structure formation in the universe substantially is also a
phenomena of structural ``surface" growth: an initially flat or smooth 
3-dimension surface described by the Robertson-Walker metric evolved 
into a roughen one. Consequently, the cosmic gravitational clustering
should also show the scaling feature as the KPZ-like surface growths.

The scaling of a KPZ-like surface is given by
$\langle (h^2-\bar{h}^2)\rangle \propto L^{\chi}f(t/L^{z})$, where $L$ is
the length of the coarse-grained average; $t$ is time. The function $f(x)$
is $\sim x^{\chi/z}$ when $x$ is small, and $f(x) \sim $ const. when $x$
is large (Vicsek 1992). The scaling indexes $\chi$ and $z$ depend on the
indexes of the spectrum of the stochastic force or the initial 
perturbations on the surface. Because $r_{cl}$ is, in fast, the scale of 
a coarse-grained average of mass field of the universe, we have
\begin{equation}
\langle M \rangle \propto r_{cl}^{\alpha}f(t/r_{cl}^{\beta})
\end{equation}
Since the scaling indexes $\alpha$ and $\beta$ depend on the indexes of 
the power spectrum of the initial perturbations, they are model-dependent. 

 From eq.(4) one can generally conclude the existence of scaling during
quasilinear and non-linear regimes. The scaling is characterized by 
1.) for small scale $r_{cl}$
and/or longer (later) time $t$, the coarse-grained mass distributions
have scaling $\langle M \rangle \propto r_{cl}^{\alpha}$; 2.) for larger 
$r_{cl}$ and/or earlier time $t$, the mass distribution will deviate from  
the $\alpha$ scaling. Because larger $r_{cl}$ and earlier time $t$
corresponds to quasilinear regime, the deviation from $\alpha$-scaling
should be due to the difference between the scaling behaviors of the
quasilinear and non-linear regimes. The indexes $\alpha$ and $\beta$ are
initial-perturbation dependent. Thus, one can expect that the
$\alpha$-scaling and the deviation from this scaling are useful for
discriminating among models of structure formation.

These general conclusions are illustrated by some semi-analytical approaches. 
For instance, with the assumption of ``stable clustering", the index
$\gamma$ of the two-point correlation function $\xi(r) \propto r^{-\gamma}$
at the fully developed non-linear regime is found to be
$\gamma=3(n+3)/(n+5)$ (Peebles 1965). This turns to $\alpha= 3/(n+5)$
(Pandmanabhan \& Engineer 1998). This means that the $\alpha$-scaling 
of highly virialized halos is indeed dependent on the initial density
perturbations.

As for the deviation from the $\alpha$-scaling, one can refer to
the power transfer via mode coupling in quasilinear regime. It has
been found that for CDM-like spectrum, the power transfer of density
perturbations is from large
scales to small ones (e.g. Suto \& Sasaki 1991).  This is, the larger scale
perturbations relatively have
higher power in the quasilinear regime than fully developed non-linear
regime. Therefore, the index $\alpha$ in quasilinear regime will generally
be larger than that of non-linear regime. Thus, one can predict that
the index $\alpha$ should show a ``going up" on larger scales.
This conclusion can also be illustrated by spherical in-fall model. Using 
scale-invariant spherical in-fall approximation, it has been shown that an 
initial scaling of $\alpha_L$ may lead to a scaling of 
$\alpha' = 3/(4-\alpha_L)$. For
$\alpha=(3-n)/2$, we have $\alpha'= 6/(n+5)$, which is larger than the
$\alpha$ given by ``stable clustering". Therefore, the scaling index in
the quasilinear regime is larger than that of non-linear regime
(Padmanabhan 1996).
 
Despite these semi-analytical results are consistent with the scenario  of 
KPZ-like dynamical scaling in general, it is difficult to semi-analytically
calculate the accurate relation between the scaling index and various initial 
perturbations. One cannot test models by the scaling index calculated from 
assumptions like the ``stable clustering" or spherical in-fall. The studies 
of surface growth has shown that in the case of 3-D one cannot find the index
of dynamical scaling analytically (Vicsek 1992). To find the number of the 
scaling, and to test models by this scaling, numerical study is necessary.
This motivated us to investigate the dynamical scaling of the gravitational
clustering numerically.

\section{Mass functions of clusters and its scaling}

\subsection{Multi-scale identifications of clusters}

In order to study the mass functions of clusters at different scales,
we have performed N-body simulations with the P$^3$M code (Jing \& Fang
1994; Jing et al. 1995; Wu et al. 1997b) for models of the SCDM, LCDM and
OCDM. The cosmological parameters ($\Omega_M, \Omega_{\Lambda},h,\sigma_8$) 
are taken to be (1.0,0.0,0.5,0.62), (0.3,0.7,0.75,1.0), (0.3,0.0,0.75,1.0) 
for the SCDM, LCDM and OCDM, respectively. The models with these
parameters seem to provide a good approximation to many observational
properties of the universe, especially the abundance of clusters (e.g. 
Jing \& Fang 1994; Bahcall, Fan \& Cen 1997).

The simulation parameters, including box size $L$, number of  particles 
$N_p$ and the effective force resolution $\eta$, are chosen to be 
($L$,$N_p$,$\eta)=(310 h^{-1}$ Mpc, $64^3,0.24 h^{-1}$ Mpc). We have run
8 realizations for each model. A particle has mass of 
$3.14\times 10^{13}\Omega_M h^{-1}$M$_{\odot}$, which is small enough to
resolve reliably the rich clusters with $M > 3.0 \times 10^{14} h^{-1}$
M$_{\odot}$.

To effectively identify clusters with different comoving radius, we haven't
employed the traditional friends-of-friends algorithm, but instead developed
an algorithm based on the discrete wavelet transformation (DWT) (see \S A1 of 
Appendix). 
The details of the DWT identification of clusters will be reported separately 
(Xu, Fang \& Deng, 
1998). Briefly, we first describe the distribution of the particles by a 3-D 
matrix, and then do fast 3-D {\it Daubechies 4} DWT and the reversed 
transformations to find the wavelet function coefficients (WFCs) and scaling 
function coefficients (SFCs) on various scales. For each scale, the cells with 
SFCs larger than those of the random sample by a given statistical significance 
are picked up as cluster candidates. Around each of the candidates, we place 
a $6^3$ grid with the size of cluster diameter and search for the accurate 
center. The cluster center is taken as the position with largest mass
surrounded.
The mass is measured by counting the particles within a sphere of radius $r_{cl}$,
the volume of which is the same as the cells. Whenever two clusters are
closer than 2$r_{cl}$, the cluster with smaller $M$ is deleted from the list. We
iterate the above steps for particles which have not been listed as cluster
members until no further clusters are found. We will call clusters identified
by radius $r_{cl}$ as $r_{cl}$-clusters. Since the DWT technique treats the
identification at different $r_{cl}$ in a uniform way, it is suitable to
study the $r_{cl}$-dependence of clusters.

Fig. 1 shows the derived cluster mass function $n(>M,r_{cl})$, which is
the number density of clusters with mass larger than $M$ within radius $r_{cl}$.
For all the three models the mass functions of $r_{cl} =1.5 h^{-1}$Mpc 
clusters given by
the DWT method are found to be in good agreement with those derived from
friends-of-friends identification (Jing \& Fang 1994). At higher $z$, these
two methods, the friends-of-friends and DWT, also provide the same mass 
functions for
$r_{cl} =1.5 h^{-1}$Mpc clusters. It turns out that the DWT technique of
identifying clusters is indeed reliable. This algorithm can also be applied to
real samples of galaxy distribution. Since the number densities of galaxies in
real samples are much less than those of particles of N-body
simulation samples,
sampling error may lead to some false identification of poor clusters.
However, for rich clusters, the effect of sampling error is small.

Since in our identified sample each cluster is characterized by two
parameters $M$ and $r_{cl}$, it is inconvenient to define richness by mass alone.
Instead, the relative richness can be defined by number density. This is, clusters
with mass $M_1$ and scale $r_1$ are considered to have the same richness
as clusters with $M_2$ and $r_2$ if $n(>M_1, r_1) = n(>M_2, r_2)$. In this
paper, the number density $n$ is expressed as $1(d)^{-3}$, where $d$ is the mean
separation of the clusters. Therefore, clusters with
$n(>M, r_{cl}) \leq 1 (50 h^{-1}$ Mpc)$^{-3}$ correspond to rich clusters of
$M > 5.5 \times 10^{14} h^{-1}$ M$_{\odot}$ on scale $r_{cl}=1.5 h^{-1}$ Mpc
for SCDM at z=0.

\subsection{The scaling of mass functions}

It can be easily seen from Fig. 1 that all the mass functions on various 
scales have a
similar shape. This is, if the 1.5 $h^{-1}$ Mpc mass functions are shifted
 horizontally
along $M$-axis, they can approximately coincide with the mass functions of 
$r_{cl}=3, 6$ and
12 $h^{-1}$ Mpc clusters, respectively. Especially, a good match is found in 
the range of abundances lower than, or richness higher than, 
1 (50 $h^{-1}$ Mpc)$^{-3}$.

This similarity can indeed be described by a scale invariance of the abundance 
mentioned in \S 2
\begin{equation}
n[>(r_{cl}/r_0)^{\alpha}M, r_{cl}] \simeq n(>M,r_0)
\end{equation}
where $\alpha$ is the index of the scaling. Eq.(5) indicates that the number
density of halos (clusters) having masses $>M$ within radius $r_0$ is the 
same as that of halos having masses $> (r_{cl}/r_0)^{\alpha}M$ within $r_{cl}$.
We have tested eq.(5) by calculating $\alpha$ for $n(>M,r_{cl})$ in the
abundance range of
$n(>M,r_{cl}) < 1 (50  h^{-1}$ Mpc)$^{-3}$, and found that $\alpha$ remains
roughly to be a constant within $5-10\%$ in the radius range
$r_{cl} \sim 1 - 6 h^{-1}$ Mpc when $r_0 = 1.5 h^{-1}$ Mpc.

The scale-invariance is conveniently expressed by a mass-radius scaling
which is the solution of the following equation:
\begin{equation}
n[>M(r_{cl}),r_{cl}]=n[>M(1.5),1.5]
\end{equation}
where 1.5 denotes $r_{cl}=1.5 h^{-1}$ Mpc. If eq.(5) is correct, we will have
\begin{equation}
M(r_{cl})=\left (\frac{r_{cl}}{1.5} \right )^{\alpha}M(1.5)
\end{equation}
(In this paper, we denote M from scaling by $M(r_{cl})$, the mass within r
by $M(r)$.)

In Fig. 2, we plot $\log [M(r_{cl})/M(1.5)]$ against $\log r_{cl}$ from
the solution
of eq.(6) for the three models, in which the ``richness" is taken to be 
$n[>M(1.5),1.5]=1 (90 h^{-1}$ Mpc)$^{-3}$. 
The loci of $\log [M(r_{cl})/M(1.5)]$ vs. $\log r_{cl}$ can be fairly
well
approximated by straight lines in the range of $1.5 < r_{cl} < 6 h^{-1}$ Mpc.
Especially, for the LCDM the relation of 
$\log [M(r_{cl})/M(1.5)]$ - $\log r_{cl}$ is quite straight in this range.
For SCDM and OCDM they also do not deviate too much from straight lines.
Therefore, the slopes of these lines, or  $\alpha$'s, nearly remain constant 
in the radius range $1.5 < r_{cl} < 6 h^{-1}$ Mpc and ``richness" range
$n[>M(r_{cl}), r_{cl}] \leq 1 (50 h^{-1}$ Mpc)$^{-3}$. The straight line
fitting yields $\alpha \approx 0.60$ for LCDM and OCDM, while
$\alpha \approx 0.80$ for SCDM.

Fig. 2 shows that all curves $\log [M(r_{cl})/M(1.5)]$-$\log r_{cl}$ are
going up around $r_{cl} \sim 6 h^{-1}$ Mpc. In other words, the index
$\alpha$ becomes larger in the range of $r_{cl} > 6 h^{-1}$ Mpc. This is
qualitatively in good agreement with the behavior of the 
scaling predicted in
last section. As the gravitation clustering on scales larger than 
$6 h^{-1}$ Mpc probably still partially remains of infall evolution, the 
scaling index is affected by the quasilinear evolution, and higher than 
that on scales less than $6 h^{-1}$Mpc. Indeed, on such large scales, more 
and more $(M,r_{cl})$-identified halos don't have dense virialized cores in
their centers, but are with complicated and irregular shapes, one of which 
is illustrated in the Fig. 1 of Jing and Fang (1994). We also find
that the  ``going up" behavior of the scaling index $\alpha$ is more 
significant at the quasilinear regime at earlier times. A similar result 
can also be seen in the comparison of universal profile with N-body
 simulation on larger scale (Diaferio \& Geller, 1997).

\subsection{The ``universal" density profile}

It has been proposed that the mass density profiles of regular
clusters may universally be fitted by $M(r) \propto \ln (1+r/a)-r/(r+a)$,
where $a$ is proportional to the virialized radius of clusters.
Both $r$ and $a$ are in unit of $r_{200}$, the radius where the average
interior density is 200 times higher than the critical density of the
universe (Navarro, Frenk \& White 1996). In the range of $r \geq a$, the
universal mass profile can be approximated as a power law
\begin{equation} 
M(r) \propto r^{\alpha_{pro}}
\end{equation}
Obviously, the index $\alpha_{pro}$ depends on the fitting of developed
virialized core with $a$ (Carlberg et al. 1997b).

Eq.(8) looks very similar to eq.(7). However, we should not identify
the index $\alpha_{pro}$ with $\alpha$, because $r_{cl}$ in eq.(7) is the 
scale used
to identify clusters, while $r$ in eq.(8) is the radius of each cluster. 
$\alpha_{pro}$
will be the same as $\alpha$ only if clusters identified by different
$r_{cl}$ have the same mass density profile. Fig. 3 shows, however, a
systematic difference of mass density profile of clusters with different
$r_{cl}$. Moreover, the larger the scale, the larger the systematic
difference. This is because the clusters identified by large $r_{cl}$ are
still partially in the quasilinear regime for which the halos are
pre-virialized. Halos, which lack a virialized core in their centers, 
cannot be described by the universal density profile.

It should be pointed out that an original meaning of the ``universal"
is the independence of profile on the initial condition, i.e., $\alpha_{pro}$
is a universal number regardless the parameters of initial spectrum like the
index $n$. 
Namely, all initial parameters are forgotten in the late time (non-linear
regime) clustering. Theoretically, it is far from clear of the condition
under which the late time profiles do not remember the initial condition. At
least, the assumption of the stable clustering cannot co-exist with an
initial-condition-independent profile (Pandmanabhan \& Engineer 1998).
Fig. 3 shows that the mean profiles $M(r)$ of clusters are significantly
different for different models. Namely, the late time profiles do remember
the initial condition.

\section{Scaling invariance of IRAS clusters}

\subsection{Multi-scale identified clusters from IRAS}

We can directly test the scale invariance via eqs.(6) and (7). To do this,
we have applied the multi-scale DWT method to identify galaxy clusters in
the redshift survey of the Infrared Astronomical Satellite (IRAS) galaxies
with flux limit of $f_{60}\ge1.2Jy$ (Fisher et al. 1995b). The IRAS surveys
are uniform and complete down to galactic latitudes $\pm 5^{o}$ from the
galactic plane. The 1.2 Jy IRAS survey consists of 5313 galaxies, which
cover 87.6\% of sky, and with $12.34\%$ of the sky belonging to the so-called
``excluded zones".

We fill up the ``excluded zones" and holes with a randomly distributed
galaxies, which are generated with the same mean number density, and the same
radial (redshift) selection function as other areas. This treatment may lead
to a little underestimate of the number density of clusters. However, in
terms of the ratio between abundances of clusters on different scales the
effect of ``excluded zones" is small. Moreover, random data, like Poisson
distributions, may give rise to false statistics on order higher than second,
but this problem will not be met in the statistics of abundance.

As usual, to reduce the effect of radial selection function, we divide the
sample into a series of shells with thickness $\sim$ 2500 km s$^{-1}$ in
redshift space, and measure the number count functions $n(>N_g, r_{cl})$ in
three shells of $v$: [2500-5000], [5000-7500] and
[7500-10000] km s$^{-1}$. There are only few galaxies
beyond 10000 km s$^{-1}$. They are not suitable for studying the scaling of
the cluster abundance.

With these redshift shells $z_i$ the density field $\delta\rho({\bf s},z_i)$ 
can be reconstructed  by a 2-D DWT decomposition with bases
$\psi_{{\bf j,l}}({\bf s})$ as 
\begin{equation}
\delta \rho({\bf s}, z_i) = \sum_{j} \sum_{l=0}^{2^j-1}\tilde{w}_{{\bf j,l}}
\psi_{{\bf j,l}}({\bf s})
\end{equation}
where the WFCs $\tilde{w}_{{\bf j,l}}$ can be found from the galaxy distribution
of the sample (A2 Appendix).  

 Since the wavelet bases $\psi_{{\bf j,l}}$ are orthogonal and complete,
the density field $\delta \rho({\bf s}, z_i)$ is the same as descriptions
 by other orthogonal-complete decomposition, say the bases of spherical
harmonics and Bessel functions (Fisher et al. 1995a). It has been 
shown that the power spectrum detected by
the DWT is the same as that by a Fourier decomposition (Pando \& Fang 1998).
The mass density given by the DWT is also the same as that of Bessel-spherical
harmonics. We choose the DWT bases only because it is easy to
detect the $j$-dependence, or scale-dependence.

 We will use $\delta \rho({\bf s}, z_i)$ to identify clusters on
various scales $j$ by the same way developed in \S 3.1. Similar to the N-body 
simulation, the cluster center is defined as the point
around which a maximum number of galaxies are enclosed within a cylinder with
length of 3000 km s$^{-1}$ and radius equal to that of the cluster.
Whenever two clusters are closer than 2$r_{cl}$, the cluster with smaller
number of galaxies $N_g$ is deleted from the list. We iterate the steps
until no further clusters are found. The result gives a number
count function $n_{IRAS}(>N_g, r_{cl})$, which is the number density of IRAS
clusters consisting of $>N_g$ galaxies within radius $r_{cl}$.

\subsection{$N_g$-$M$ conversion}

Mass is not directly measurable for IRAS clusters. To transfer the 
number count function $n(>N_g, r_{cl})$ to mass function $n(>M,r_{cl})$,
we need a conversion between the number count and mass for the IRAS
clusters. Because the total luminosity of a $r_{1.5}$-cluster is 
proportional to its richness (mass), and the mass-luminosity ratio for 
these cluster is
independent of the total mass of the clusters (Bahcall \& Cen 1993),
it is reasonable to assume that the mass $M$ of a cluster is
proportional to the total number $N_{total}$ of the galaxies in the given
cluster, i.e.
\begin{equation}
M = A N_{total}
\end{equation}
Obviously, $A$ is only a number-mass conversion coefficient, not
the mean mass of galaxies, as the cluster mass $M$ is dominated by dark 
matter. The total number of galaxies is related to the counted galaxy by
the selection function $\phi(z)$(Fisher et al. 1995b):
\begin{equation}
 N_{total} = N_g/\phi(z)
\end{equation}
We have then
\begin{equation}
M = A * N_g/\phi(z)
\end{equation}
We calibrate the coefficient A at a fixed mass, i.e., at a given
abundance $n \approx 10^{-5} h^3$ Mpc$^{-3}$ through the
equation:
\begin{equation}
n_{IRAS}(>N_g, 1.5)= n_{BC}(>M,1.5)
\end{equation}
where subscript $BC$ means the mass function of Bahcall and Cen
(1992), and $n_{IRAS}(>N_g, 1.5)$ is from the shell of 
[5000-7500] kms$^{-1}$. We solve for $N_g$ and $M$.
 Using this pair of $N_g$ and $M$, we derive 
the coefficient  A from eq.(12).

With this $A$, one can transfer the number count function 
$n_{IRAS}(>N_g, r_{cl})$ into mass function $n_{IRAS}(>M,r_{cl})$ for
the entire range of $N_g$. The result of $n_{IRAS}(>M,r_{cl})$ is plotted 
in Fig. 4, in which the horizontal errors of IRAS clusters are caused
by the Poisson errors of number counting of galaxies in a given 
cluster, and the vertical errors are from the Poisson errors of 
counting the clusters.

The BC mass function is also shown in Fig.4. It has been known
that SCDM model doesn't fit BC's MF while LCDM fits well(Jing \& Fang
1994). IRAS results confirm this conclusion. One can find that 
the mass function of the IRAS clusters with $r_{cl}= 1.5 h^{-1}$ Mpc 
is basically the same as the mass function of BC sample, especially, for 
clusters with richness $M >10^{14} h^{-1}$ M$_{\odot}$. Namely, 
the masses of clusters identified from the 1.2 Jy IRAS samples are 
statistically the same as those of the clusters in the BC sample. This 
result is consistent with the following fact: the IRAS galaxies trace 
the local large scale structures. It has been found that many optically
identified 
structures, including superclusters, voids, great attractor and Abell 
clusters, have been identified from density field constructed from 
IRAS data (Webster, Lahav \& Fisher 1997). Considering that the 
clusters of BC sample consist of optical and X-ray clusters, and the
IRAS galaxies are biased, containing more later type galaxies, Fig. 4 
indicates that the early-type galaxies map about the same mass field as 
late-type, despite the early-type galaxies are clustered more strongly than
late-type galaxies. This is because that in terms of second order of
statistics the segregation between the early
and late-type galaxies is almost linear, and scale-independent at
relatively large scales. Using Stromlo-APM redshift survey it has been 
shown that for second order statistics the scale-dependence of the
segregation between early and 
late-type galaxies is not large than 1 $\sigma$ in the range from 1 
to 20 $h^{-1}$ Mpc (Loveday et al. 1995, Fang, Deng \& Xia 1998).

One can further test the reasonableness of the calibration (12) by
comparing different redshift shells. With eq. (12) and the
selection function $\phi(z)$ (Fisher et al. 1995b), we found 
that the best values of $A$ for the three shells are, respectively,
$10^{12.1\pm0.2}, \ 10^{11.9\pm0.2}, \ 10^{11.7\pm0.3} \ 
h^{-1}$ M$_{\odot}$. They are indeed the same within error limit. Thus, the
 mass functions $n_{IRAS}(>M,r_{cl})$ 
from the shells of  [2500-5000] and [7500-10000] kms$^{-1}$ 
are the same as [5000-7500] km s$^{-1}$, and also the same as the BC sample.
Therefore, the number of $A$ provides a consistent $N_g-M$ conversion
for the entire sample of the 1.2 Jy IRAS galaxies. Using the conversion
of Abell richness to cluster mass: $M/N_{c}=0.6\times10^{13}h^{-1}$
M$_{\odot}$ (Bahcall \& Cen 1993),  we have
$A \equiv M/N_{total}=0.8\times10^{12}h^{-1}$
M$_{\odot}$ gives $N_{c}/N_{total}\approx 7$. It means every count of the
1.2 Jy IRAS galaxies (after selection-function correction) corresponds to
about 7 times Abell count of optical galaxies in counting the mass of a
cluster.

\subsection{Scaling of IRAS clusters and models}

Because $M/L$ reaches a constant asymptotic value beyond 
$r_{cl} \sim 1$ Mpc (Bahcall, Lubin \& Dorman 1995), and there is no evidence
of significant scale dependence of bias factor of IRAS galaxies from 1 to
10 h$^{-1}$ Mpc, the $N_g$-$M$ conversion eq.(12) should be applicable on
scales $ > 1.5 h^{-1}$Mpc.
Thus we can find mass functions $n(>M, r_{cl})$ from number-count functions  
$n(>N_g, r_{cl})$ of IRAS cluster with $r_{cl} > 1.5 h^{-1}$Mpc. 

Using these IRAS $n(>M,r_{cl})$, we test the scaling by eq.(6). The solutions
of eq.(6) for both IRAS data and simulation sample are shown in Fig. 5.
The theoretical curves in Fig. 5 are similar to Fig. 2, and the richness
parameter is taken to be $1 (50 h^{-1}$ Mpc)$^{-3}$.

Since only mass ratios $M(r_{cl})/M(1.5)$ of the solutions eq.(6) are plotted 
in Fig. 5, the result doesn't depend on the value of A. The effect of linear bias 
of galaxies will also be canceled in this ratio. The errors of IRAS data in Fig. 5
are calculated from both horizontal and vertical errors of the mass function 
(Fig. 4).
Since the mass function is very steep for rich clusters, i.e. 
$|d\ln n/d\ln M| \gg 1$, an uncertainty of $\ln n$ will transfer to a relatively 
small uncertainty of $\ln M$. Because the mean number density of 1.2 Jy IRAS 
galaxies is low, the major source of errors in Fig. 5 is from Poisson errors of 
the number of galaxies in clusters.

Despite the errors are large, Fig. 5 already shows that the IRAS data is basically  
consistent with the prediction: there is a scaling invariance in the range of 
$1.5 < r_{cl} < 4.5 h^{-1}$Mpc with index $\alpha \sim 0.5$, and the scaling 
index of scaling is ``going up" on larger scales. The results of simulation
samples show that the index of scaling is model-dependent. The three panels
of Fig. 5 generally  are
in agreement with the two low mass models (LCDM and OCDM) within 1 $\sigma$,
but show a systematic disfavor of the SCDM at $\geq 1 \sigma$.
The error bars in Fig. 5 are slightly overestimated by assuming that the
Poisson errors of mass estimates between two scales are independent.
They might be about $1/\sqrt{2}$ times smaller if the two Poisson
errors are completely correlated (cloud in cloud scenario). So,
in each of the three shells the SCDM is away from the data at the level
of $\geq 1\sim 1.4 \sigma$, or disfavored at a confidence level
of $\geq 68\%-82\%$.  If all the shells are binned together,
the confidence should certainly be higher
than that of individual shell, because the Poisson errors will be smaller. 
However, it is difficult to estimate 
the influence of  the selection function.
 Moreover, if spiral galaxies were underrepresented within about 1.5
$h^{-1}$ Mpc, the true values of the scaling index should be lower than
that shown in Fig. 5, i.e. it strengthens the conclusion of disfavoring the
SCDM. Therefore, the number $68\%-82\%$ can be applied as a
underestimated confidence level.

\section{Other detections of the scaling index}

\subsection{Sample of rich clusters with mass estimation}

 Similar to eq.(2), the number density of $r_{cl}$ clusters with mass in
the range of $M$ to $M+dM$ is
\begin{equation}
N(M,r_{cl})dM = - \frac {\partial}{\partial M} n(>M,r_{cl})dM.
\end{equation}
One can then define a mean mass of $r_{cl}$ clusters with richness
$n(>M,r_{cl}) < n_0$ by
\begin{equation}
\overline{M(r_{cl})} = \int_{n(>M,r_{cl}) < n_0} M N(M,r_{cl}) dM.
\end{equation}
Using eqs.(5), (6) and (7), we have
\begin{equation}
\log \overline{M(r_{cl})} = \alpha \log r_{cl} + {\rm const}.
\end{equation}
This means, the index $\alpha$ determined by fitting eq.(16) with a
sample of mass measurements of clusters with richness $n(>M,r_{cl}) < n_0$
should be the same as that given by abundance solution of eqs.(6) and
(7). 
This prediction can be tested if we have fair samples of masses of clusters 
with richness $n(>M,r_{cl}) < n_0$ and over a given radius range of $r_{cl}$.

Despite we still lack cluster mass samples covering a large 
radius range of $r_{cl}$ and with the desired completeness, the current 
data are already able to preliminarily test the prediction eq.(16). 
For instance, it is generally believed that the weak gravitational 
lensing clusters being studied are among the 
richest clusters. Weak lensing mass estimate gives only a lower bound 
to the total cluster mass because of the unknown mean density in the so-called 
control annulus (Fahlman et al 1994). Nevertheless, the radius dependence of 
the cluster mass given by weak gravitational lensing, $M_{wl}(r_{cl})$,
is insensitive to the control field which contributes only a constant 
component to the mass distribution.  Moreover, under the isothermal 
assumption, we have $M(r_{cl})/r_{cl} \propto \sigma_1^2$, where $\sigma_1$ is the
sight-of-line velocity dispersion of the cluster galaxies. Since the 
relation $M(r_{cl})/r_{cl} \propto \sigma_1^2$ is independent of richness, a
velocity dispersion normalized mass, $M_{wl}(r_{cl})/\sigma_1^2$, appears
to be less dependent on richness. So, the cluster mass sample consisting of
$M_{wl}(r_{cl})$ and velocity dispersion measurements would be good for fitting
with eq.(16). There are 9 clusters which have both velocity dispersion 
measurements and weak gravitational lensing mass $M_{wl}$ 
estimates in the radius range of 0.15 to 2 $h^{-1}$Mpc (Wu \& Fang 1997a).
Although $M_{wl}(r_{cl})$ from the weak lensing actually corresponds to a
projected mass within $r_{cl}$, the values of $\alpha$ given by either the
projection or the 3-D masses will be roughly the same at large radius  
$r_{cl}>1 h^{-1}$ Mpc. A best fitting of this weak lensing data yields 
$\alpha=0.71\pm 0.20$. 

Another data set of rich cluster masses can be selected from X-ray 
measurements. 
The largest sample of X-ray clusters with mass estimates published to date 
is given by White, Jones \& Forman (1998), which contains 226 X-ray
cluster masses for 207 clusters derived from a deprojection of Einstein 
Observatory X-ray imaging data. Meanwhile, by extensively searching
literature  there are additional 152 X-ray determined cluster/group masses
available. These data were obtained by either the similar approach to WJF
(e.g. White \& Fabian 1995; Ettori, Fabian \& White 1998; etc.) or the 
analysis of the ROSAT PSPC X-ray observations (e.g. Pildis. et al 1995; David, 
Jones \& Forman 1995; Cirimele et al. 1997; etc.). All the available 144 
measurements of cluster masses from X-ray observations with $r_{cl}>0.5
h^{-1}$Mpc are plotted in Fig. 6.

The X-ray and optical measurements of some gravitational lensing clusters 
have shown that the gravitational lensing clusters on average correspond to 
X-ray temperature $\overline{T} \geq 7.5$ keV and velocity dispersion of 
$\sigma_1 \geq 1200$ km s$^{-1}$. Therefore, it is reasonable to select 
richest clusters by the conditions of $\overline{T} \geq 7$ keV and 
$\sigma_1 \geq 1000$ km s$^{-1}$. There are 11 measurements
satisfying these conditions. These data are plotted as star-added circles 
in Fig. 6. It yields $\alpha=0.66\pm0.40$ in the range of 
$1.5 <r_{cl}< 4 h^{-1}$Mpc. 
 
It is interesting to see from Fig. 6 that the X-ray mass distribution has a
clear upper envelope, and all the clusters selected by $\overline{T} \geq 7$ keV
and $\sigma_1 \geq 1000$ km s$^{-1}$ distribute along the envelope. 
It implies that the clusters at the envelope are among the richest clusters 
at the given $r_{cl}$. Recall that observations may easily lose less massive
clusters, but tend to pick up the bright and massive ones. This 
selection effect is severe for producing a reliable sample of less rich 
clusters, but benefit to the completeness of sample of rich clusters.
Therefore, one can employ the envelope clusters to fit with eq.(16). 
To this end, we have binned the data set of cluster masses into 8 
logarithmic intervals according to radius from $r_{cl}= 0.5 - 4.0 h^{-1}$Mpc.
Within each bin, the mean value of the second and the third largest clusters
is used as the envelope value. Two of the 8 bins having less than 2 clusters
are not considered as reliable measurements of the envelope and are thus not 
used.
All envelope values are shown in Fig. 6, in which the vertical error bars are
the mass difference between the second and third masses of the most massive
clusters. The least-square fitting gives $\alpha=0.52\pm0.25$ in the
radius range of 0.5 to 3 $h^{-1}$Mpc. To reduce the possible effect of the 
bin size selection, we re-calculated $\alpha$ using different bin sizes. It 
turns out that differences among these results are not larger than $1 \sigma$.
 
All the detections of eq.(16) with independent ensembles of rich cluster
mass estimates have yielded the same number of $\alpha$ within error limits. 
For clusters with ``richness"
$n(>M, r_{cl}) \leq 1 (50 h^{-1}$ Mpc)$^{-3}$ in 
the range of $1 < r_{cl} < 4 h^{-1}$ Mpc the mean value of $\alpha$ is
$\simeq 0.63 \pm 0.10$. These values of $\alpha$ basically agree with the 
$\alpha$ detected by the abundance of IRAS clusters (\S 4.3). Therefore, current 
data are in good consistence with the scaling invariance of rich cluster 
abundance.

\subsection{Deviation from a constant $\alpha$}

 As has been discussed in \S 2, the scaling index $\alpha$ remains as a constant 
on smaller scales and later time, but will show ``going up" on larger
scales and early time on which the quasilinear evolution still plays important role.
Therefore,  one can expect that the ``going up" behavior will be more significant 
at higher redshifts. An effective measure of the ``going up" is the abundance 
ratio $n[>(r_{cl}/r_0)^{\alpha}M,r_{cl}]/n(>M,r_0)$ at higher redshifts. If the 
scaling invariance of abundance perfectly holds with a constant index $\alpha$, 
we have 
\begin{equation}
\frac{n[>(r_{cl}/r_0)^{\alpha}M,r_{cl}]}{n(>M,r_0)} =1
\end{equation}
Therefore, the radius $r$ beyond which this ratio no longer remains equal to 
unity is a measure of the importance of the quasilinear evolution.  

Fig. 7 plots the ratio $n[>(r_{cl}/r_0)^{\alpha}M,r_{cl}]/n(>M,r_0)$ of simulated
samples with parameters $\alpha= 0.63$, $r_0 = 1.5 h^{-1}$ Mpc and
$M=5.5 \times 10^{14} h^{-1}$M$_{\odot}$. Fig. 7 shows that all the
abundance ratios become larger than 1 on large scales. Namely the real values 
of $\alpha$ for the simulated samples are larger than the assumed 0.63 
on larger scales. This is the ``going up", indicating the deviation from 
constant $\alpha$.

Fig. 7 shows that this ``going up" behavior is significantly different for 
different models. The SCDM curves are much more quickly and strongly
``going up"  than LCDM and OCDM. This is because clusters
in the SCDM formed later than in
the LCDM and OCDM. In the latter case, the ratio
$n[>(r_{cl}/r_0)^{\alpha}M,r_{cl}]/n(>M,r_0)$ remains equal to about unity
in the range of $r_{cl} = 1.5 - 6 h^{-1}$ Mpc and $z \leq 0.8$, while for
the former the corresponding radius range is much smaller. We have
re-calculated Fig. 7
by changing the value of $\alpha$. The results show that in the range 
$\alpha \sim 0.43 - 0.77$, the OCDM and LCDM always have a larger radius
range  of the invariance than SCDM. Therefore, the radius range in which the
abundance scaling invariance holds is effective for discriminating models.

The fitting done in \S 5.1 has shown that a constant $\alpha$ 
of 0.52 - 0.70 in the radius $r_{cl}$ from 1 to 4 $h^{-1}$ Mpc is consistent with
all data. Moreover, considering most gravitational lensing and X-ray clusters 
used in \S 5.1 are of moderate redshift, the numbers of $\alpha \sim 0.52 - 0.70$ 
and $1< r_{cl} < 4 h^{-1}$ Mpc are actually real for moderate redshift.
Thus, the SCDM will be in difficulty with the measurements of scaling 
invariance. It should be pointed out that the radius range
$1 < r_{cl} < 4 h^{-1}$ Mpc in \S 5.1 is for the physical scale
of relevant clusters. A physical radius range 1 - 4 $h^{-1}$ Mpc corresponds 
to a comoving range of 1.5 - 6 $h^{-1}$ Mpc at redshift $\sim 0.5$. Therefore, 
this test is more robust at higher redshift. 

It may be difficult to directly test the scaling
at high redshifts, because mass determination is often 
limited by the luminosity(or surface brightness) detection
threshold of the survey.
In this case we may turn to the correlation function. The scaling
gives testable prediction on the correlation lengths of clusters
on different scales. This will be investigated in Xu, Fang \& Deng (1998).

\section{Conclusions}

We have studied the scaling invariance of abundance of rich clusters. 
Both the N-body simulation and the available observational data have 
confirmed the existence of the scaling invariance. This scaling is 
characterized by index $\alpha$ which can be determined by
the abundances of clusters on different scale $r_{cl}$, or by fitting 
mass-radius
relation of clusters. The scaling gains a further support from the 
following result: the  $\alpha$ given by X-ray and gravitational lensing 
mass estimates is the same as that from the IRAS cluster abundance. 
Despite the significance level of current results is not high, it is worth 
of revealing the general behavior of the scaling of cosmic clustering in 
different evolutionary stages. It can already be employed for 
discrimination among models.

The LCDM and OCDM abundances can always have a scaling in a larger
scale range ($\sim 1 - 6 h^{-1}$ Mpc) with  $\alpha \sim 0.5 - 0.7$ 
for $z \leq 0.8$. While the SCDM can only provide a smaller scaling range
($\sim 1 - 3 h^{-1}$ Mpc) with $\alpha < \sim 0.8$ and $z<0.8$. If 
$\alpha$ is allowed to be $\geq 0.80$, the SCDM can provide a scaling
in the range of 1 - 4 $h^{-1}$ Mpc for $z \leq 0.8$. However, the SCDM is 
difficult to fit $\alpha \simeq 0.50 - 0.70$ and 
$r_{cl} \sim 1 - 4 h^{-1}$ Mpc simultaneously. This result may cause some
 troubles for the SCDM.

In general, the mass density profiles of clusters in the low density models
are steeper than that of the corresponding clusters in higher $\Omega_M$ models
(Jing et al. 1995). In other words, the index $\alpha$ for the $\Omega_M=1$
model is always larger than that in a low density model ($\Omega_M <1$).
Therefore, the current result of the mass-radius scaling seems to favor
models with a lower mass density.

\acknowledgments

WX thanks the World Laboratory for a scholarship. XPW was  
supported by the National Science Foundation of China and
the National Science Council of Taiwan.

\appendix
\section{The DWT decomposition and reconstruction}

\subsection{The discrete wavelet transform}

Let us briefly introduce the DWT analysis of large scale structures, for 
the details referring to (Fang \& Pando 1997; Pando \& Fang 1996, 1998,
and references therein).
We consider here a 1-D mass density distribution  $\rho(x)$ 
or contrast $\delta(x)=[\rho(x)-\bar{\rho}]/\bar{\rho}$, which are 
mathematically random fields over a spatial range $0 \leq x \leq L$. It 
is not difficult to extend all results developed in this section into 
2-D and 3-D because the DWT bases for higher dimension can be constructed 
by a direct product of 1-D bases.

Like the Fourier expansion of the field $\delta(x)$, the DWT expansion of 
the field $\delta(x)$ is given by
\begin{equation}
\delta(x) = \sum_{j=0}^{\infty} \sum_{l= 0}^{2^j -1}
  \tilde{w}_{j,l} \psi_{j,l}(x)
\end{equation}
where $\psi_{j,l}(x)$, $j=0,1,...$, $l=0...2^j-1$ are the bases of the DWT.
Because these bases are orthogonal and complete, the wavelet function
coefficient (WFC), $\tilde{w}_{j,l}$, is computed by
\begin{equation}
\tilde{w}_{j,l}= \int \delta(x) \psi_{j,l}(x)dx.
\end{equation}

The wavelet transform bases $\psi_{j,l}(x)$ are generated from the 
basic wavelet $\psi(x/L)$ by a dilation, $2^j$, and a translation 
$l$, i.e. 
\begin{equation}
\psi_{j,l}(x) =\left( \frac{2^j}{L} \right)^{1/2} \psi(2^jx/L-l).
\end{equation}
The basic wavelet $\psi$ is designed to be continuous, admissible 
and localized. Unlike the Fourier bases $\exp (i2\pi nx/L)$, which are 
non-local in 
physical space, the wavelet bases $\psi_{j,l}(x)$ are localized in 
both physical space and  Fourier (scale) space. In physical space, 
$\psi_{j,l}(x)$ is centered at position $lL/2^j$, and in Fourier 
space, it is centered at wavenumber $2\pi \times 2^j/L$. Therefore,
the DWT decomposes the density fluctuating field 
$\delta(x)$ into domains ${j,l}$ in phase space, and for each basis the
corresponding area in the phase space is as small as that allowed by the
uncertainty principle. WFC $\tilde{w}_{j,l}$ and its intensity
$|\tilde{w}_{j,l}|^2$ describe, respectively,
the fluctuation of density and its power on scale $L/2^j$ at position
$lL/2^j$ (Pando \& Fang 1998).

\subsection{Reconstruction of density field}

Using the completeness of the DWT basis, one can reconstruct the original
density field from the coefficient $\tilde{w}_{j,l}$. To
achieve this, DWT analysis employs another set of functions consisting of
the so-called scaling functions, $\phi_{j,l}$, which are generated from
the basic scaling $\phi(x/L)$ by a dilation, $2^j$, and a translation $l$,
i.e.
\begin{equation}
\phi_{j,l}(x) = \left( \frac{2^j}{L} \right )^{1/2} \phi(2^jx/L - l).
\end{equation}
The basic scaling $\phi$ is essentially a window function with width
$x/L=1$. Thus, the scaling functions $\phi_{j,l}(x)$ are also windows, 
but with width $(1/2^j)L$, and centered at $lL/2^j$.
The scaling function $\phi_{j,l}(x)$ are orthogonal with respect to 
the index $l$, but not for $j$. This is a common property of windows, which 
can be orthogonal in physical space, but not in Fourier space. 

 For Daubechies wavelets, the basic wavelet and the basic scaling are
related by recursive equations as (Daubechies 1992) 
\begin{equation}
\begin{array}{ll}
\phi(x/L) & = \sum_l a_{l} \phi(2x/L-l) \\
\psi(x/L) & = \sum_{l} b_{l} \phi(2x/L + l)
\end{array}
\end{equation}
where coefficients $a_l$ and $b_l$ are different for different wavelet. 
In this paper, we use the Daubechies 4 wavelet (D4), for which
$a_0=(1+\sqrt 3)/4, \ a_1=(3+\sqrt 3)/4, \ a_2=(3-\sqrt 3)/4, \
a_3=(1-\sqrt 3)/4$. 

 From Eq.(A5), one can show that the scaling functions $\phi_{j,l}(x)$ are
always orthogonal to the wavelet bases $\psi_{j',l'}(x)$ if $j \leq j'$, i.e.
\begin{equation}
\int \phi_{j,l}(x)\psi_{j',l'}(x)dx = 0, 
\hspace{2cm} {\rm for \ \ \ } j \leq j'.
\end{equation}
Therefore, $\phi_{j,l}(x)$ can be expressed by
$\psi_{j',l'}(x)$ as
\begin{equation}
\phi_{j,l}(x) = \sum_{j'=0}^{\infty}\sum_{l'=0}^{2^{j'}-1}
c_{jl;j'l'} \psi_{j',l'}(x) =
\sum_{j'=0}^{j-1}\sum_{l'=0}^{2^{j'}-1}
c_{jl;j'l'} \psi_{j',l'}(x).
\end{equation}
The coefficients $c_{jl;j'l'}= \int \phi_{j,l}(x)\psi_{j',l'}(x)dx$
can be determined by $a_l$ and $b_l$.

Using $\phi_{j,l}(x)$, we construct a density field on
scale $j$ as
\begin{equation}
\rho^{j}(x) = \sum_{l=0}^{2^{j}-1} w_{j,l} \phi_{j,l}(x),
\end{equation}
where $w_{j,l}$ is called the scaling function coefficient (SFC) given by
\begin{equation}
w_{j,l}= \int_0^{L} \rho(x) \phi_{j,l}(x)dx.
\end{equation}
Since the scaling function 
$\phi_{j,l}(x)$ is window-like, the coefficient $w_{j,l}$ is actually a 
``count-in-cell" in a window on scale $j$ at position $l$.

Using Eqs.(A1), (A7), (A8) and (A9), one can find
\begin{equation}
\rho^j(x) = \bar{\rho}\sum_{j'=0}^{j-1} \sum_{l'= 0}^{2^{j'} -1}
  \tilde{w}_{j',l'} \psi_{j',l'}(x) + \bar{\rho}.
\end{equation}
Namely, $\rho^j(x)$ contains all terms of density fluctuations
$\tilde{w}_{j',l'}\psi_{j',l'}(x)$ of $j' <j$, but not terms 
of $j' \geq j$.  From Eqs.(A1) and (A10), we have
\begin{equation}
\rho(x)=\rho^j(x) + \bar{\rho}
\sum_{j'=j}^{\infty} \sum_{l'= 0}^{2^{j'} -1}
  \tilde{w}_{j',l'} \psi_{j',l'}(x).
\end{equation}
One can also define the smoothed density contrasts on scale $j$ 
to be 
\begin{equation}
\delta^j(x) \equiv \frac{\rho^j(x)-\bar{\rho}}{\bar{\rho}}
= \sum_{j'=0}^{j-1} \sum_{l'= 0}^{2^{j'} -1}
  \tilde{w}_{j',l'} \psi_{j',l'}(x).
\end{equation}
Eqs.(A11) and (A12) show that $\rho^j(x)$ and $\delta^i(x)$ are smoothed
density fields on scale $j$. One can construct the density field
$\rho^j(x)$ or $\delta^j(x)$ on finer and finer scales by WFCs
$\tilde{w}_{j,l}$ till to the precision of the original field. Since the
sets of bases
$\psi_{j,l}$ and $\phi_{j,l}$ are complete, the original field can be 
reconstructed without lost information.

\figcaption{Mass functions of clusters identified with radii $r_{cl}=$ 1.5,
3.0, 6.0 and 12 h$^{-1}$Mpc for SCDM, LCDM and OCDM models at $z=0$.
$n(>M,r_{cl})$ is the number density of clusters with mass larger than $M$
within radius $r_{cl}$. $M$ is in unit of $h^{-1}$ M$_{\odot}$. The observed
data are for clusters with radius $r_{cl}= 1.5 h^{-1}$Mpc (Bahcall \& Cen
1992). 
\label{fig1}}

\figcaption{
Mass-radius scaling of clusters given by the solution of abundance 
equation (6). The ``richness" is taken to be
$n= 1 (90 h^{-1}$ Mpc$)^{-3}$. The thin, dashed, and thick lines are for
SCDM, LCDM and OCDM, respectively. 
\label{fig2}}

\figcaption{
The mean mass-radius relation $M(r_{cl})$ of clusters with different
identification scales $r_{cl}$. The ``richness" for all $r_{cl}$ clusters is
$n= 1 (90 h^{-1}$ Mpc$)^{-3}$. The dashed and solid lines are for SCDM and 
OCDM, respectively.
\label{fig3}}

\figcaption{Mass function of $r_{cl}=1.5 h^{-1}$Mpc IRAS clusters (circles with 
error bars) in the shell [5000,7500]km s$^{-1}$. The stars are for the mass function
of Bahcall \& Cen (1992). The dashed line shows the mass function by FOFs method of 
Jing \& Fang (1994). The thick line shows the mass function by DWT method averaged 
over 5-realizations of our 310$h^{-1}$Mpc box simulations. The thin line 
shows the mass function
by DWT method from 1 realization of 155$h^{-1}$Mpc box simulation.
\label{fig4}}

\figcaption{Mass-radius scaling of clusters given by the solution of
abundance equation (2).  The thick, thin, and dashed  lines are for
LCDM, OCDM and SCDM, respectively. The data with error bars come from the
clusters identified from IRAS galaxies. The richness parameter is taken 
to be  $1 (50 h^{-1}$ Mpc)$^{-3}$.
\label{fig5}}

\figcaption{All the available 144 measurements (circles) of cluster masses from
X-ray observations with $r_{cl}>0.5 h^{-1}$Mpc.
 The 11 richest clusters($T>7$ keV and 
$\sigma>1000$kms$^{-1}$) with $r_{cl}>1.5 h^{-1}$Mpc are marked additionally
with stars.
Squares with error bars are the data of upper envelope clusters.
The horizontal error bars showing the widths of radius bins, and the vertical 
error bars being the difference between the masses of the second and the third 
most massive clusters within each bin. The solid line is an equal-weight 
least square fit to the envelope data.
\label{fig6}}

\figcaption{$n[>M_{th}(r_{cl}),r_{cl}]/n_{1.5}$ vs. $r_{cl}$ in the range of 
$r_{cl}= 1.5 - 12
h^{-1}$Mpc at redshifts $z \sim $ 0.2, 0.5 and 0.8, for SCDM, LCDM and 
OCDM. Here $M_{th}(r_{cl}) = (r_{cl}/r_{1.5})^{\alpha} M_{1.5}$ and
$n_{1.5}=n(>M_{1.5},1.5)$. The parameters are taken to be 
$M_{1.5}= 5.5 \times 10^{14} h^{-1}$ M$_{\odot}$, 
$r_{1.5}= 1.5 h^{-1}$Mpc, and $\alpha= 0.63$. The dot-dashed horizontal 
lines are for perfect scale invariance with $\alpha = 0.63$. All curves for 
simulated samples show the ``going up" of $\alpha$ on large scales.
\label{fig7}}



\begin{references}
\reference{} Bahcall, N.A. \& Cen, R.Y. 1992, \apjl, 398, L81 (BC)

\reference{} Bahcall, N. A., Cen, R. Y. 1993, \apjl, 407, L49

\reference{} Bahcall, N.A., Fan, X. \& Cen, R.Y. 1997, \apjl, 485, L53

\reference{} Bahcall, N.A., Lubin, L. M. \& Dorman, V. 1995, \apjl, 447,
L81

\reference{} Barab\'asi, A.-L. \& Stanley, H.E. 1995, {\it Fractal Concepts in
     Surface Growth}, (Cambrifdge Univ. Press)

\reference{} Bennett, C.L. et al, 1996, \apjl, 464, L1.

\reference{} Berera, A. \& Fang, L.Z. 1994, Phys. Rev. Lett., 72, 458

\reference{} Carlberg, R.G., Morris, S.M., Yee, H.K.C. \&
	 Ellingson, E. 1997a, \apjl, 479, L19

\reference{} Carlberg, R.G., Yee, H.K.C. \&  Ellingson, E. 1997b, 
     \apj, 478, 462

\reference{} Cirimele, G., Nesci, R. \& Trevese, D. 1997, \apj, 475, 11

\reference{} David, L. P., Jones, C. \& Forman, W. 1995, \apj, 445, 578

\reference{} Diaferio, A. \& Geller, M. 1997, \apj, 481, 633

\reference{}Daubechies,I., 1992, {\it Ten Lectures on Wavelets}, 
(Philadelphia, SIAM)


\reference{} Eke, V.R., Cole, S., Frenk, C.S. \& Navarro, J.F. 1996,
	\mnras, 281, 703

\reference{} Ettori, S., Fabian, A. C. \& White, D. A. 1998, \mnras, in
press

\reference{} Fahlman, G., Kaiser, N., Squires, G., \& Woods, D.
            1994, \apj, 437, 56

\reference{} Fang, L. Z. \& Pando, J. 1997, Proceedings of the Erice
	School ``Astrofundamental Physics", eds. Sanchez N. \&
	Zichichi A., (Singapore, World Scientific), 616

\reference{} Fang, L.Z., Deng, Z.G. \& Xia, X.Y. 1998, \apj, in press

\reference{} Fisher, K.B., Lahav, O., Hoffman, Y., Lynden-Bell, D. \&
    Zaroubi, S. 1995a, \mnras, 272, 885

\reference{} Fisher, K. B., Huchra, J.P., Strauss, M.A., Davis, M., 
  Yahil, A. \& Schlegel, D. 1995b, \apjs, 100, 69

\reference{} Jing, Y.P., Mo, H.J., B\"orner, G \&  Fang, L.Z. 1993,
   \apj, 411, 450

\reference{} Jing, Y.P. \& Fang, L.Z. 1994, \apj, 432, 438 

\reference{} Jing, Y.P., Mo, H.J., B\"orner, G. \& Fang, L. Z. 1995,
		\mnras, 276, 417

\reference{} Kardar, M, Parisi, G \& Zhang, Y.C. 1986, Phys. Rev. Lett.
              56, 342.

\reference{} Loveday, L., Maddox, S.J., Efstathiou, G. \& Peterson, B.A.
             1995, \apj, 442, 457

\reference{} Luppino, G.A.. \& Gioia, I.M. 1995, \apj, 445, L77

\reference{} Navarro, J.F., Frenk, C.S. \& White, S.D.M. 1996, \apj,
             462, 563
	
\reference{} Oukbir, J. \& Blanchard, A. 1997, A\&A, 317, 10

\reference{} Padmanabhan, T. \& Enginer, S. 1998, \apj, 493, 509

\reference{} Padmanabhan, T. 1996, \mnras, 278, L29

\reference{} Pando, J. \& Fang, L.Z. 1996, \apj, 459, 1

\reference{} Pando, J. \& Fang, L.Z. 1998, Phys. Rev. E57, 3593

\reference{} Pando, J., Greiner, M., Lipa, P. \& Fang, L.Z. 1998, \apj, 
               496, 9

\reference{} Peebles, P.J.E. 1965, \apj, 142, 1317

\reference{} Peebles, P.J.E. 1980, Large Scale Structure of the Universe,
  (Princeton, Princeton Univ. Press), \S 73

\reference{} Pildis, R. A., Bregman, J. N., \& Evrard, A. E. 1995, \apj,
443, 514

\reference{} Suto, Y. \& Sasaki, M. 1991, Phys. Rev. Lett. 66, 264

\reference{} Viana, P. P. \& Liddle, A. R. 1996, \mnras, 281, 323

\reference{} Vicsek, T. 1992, Fractal Growth Phenomena, (Singapore, World 
         Scientific), Chap 7

\reference{} Webster, M., Lahav, O., \& Fisher, K. 1997, \mnras, 287,
425

\reference{} White, D. A. \& Fabian, A. C. 1995, \mnras, 273, 72

\reference{} White, D. A., Jones, C., \& Forman, W. 1998, \mnras, in
press (WJF)

\reference{} Wu, X.P. \& Fang, L.Z. 1996, \apjl, 467, L45

\reference{} Wu, X.P. \& Fang, L.Z. 1997a, \apj, 483, 62

\reference{} Wu, X.P., Fang, L.Z., Chiueh, T., \& Xue, Y. 1998, in
	preparation

\reference{} Wu, X.P., Zhu, X.H., Jing, Y.P. \& Fang, L.Z. 1997b, \apj, 
             488, 557
 
\reference{} Xu, W., Fang, L.Z., \& Deng, Z. G. 1998, in preparation

\end{references}
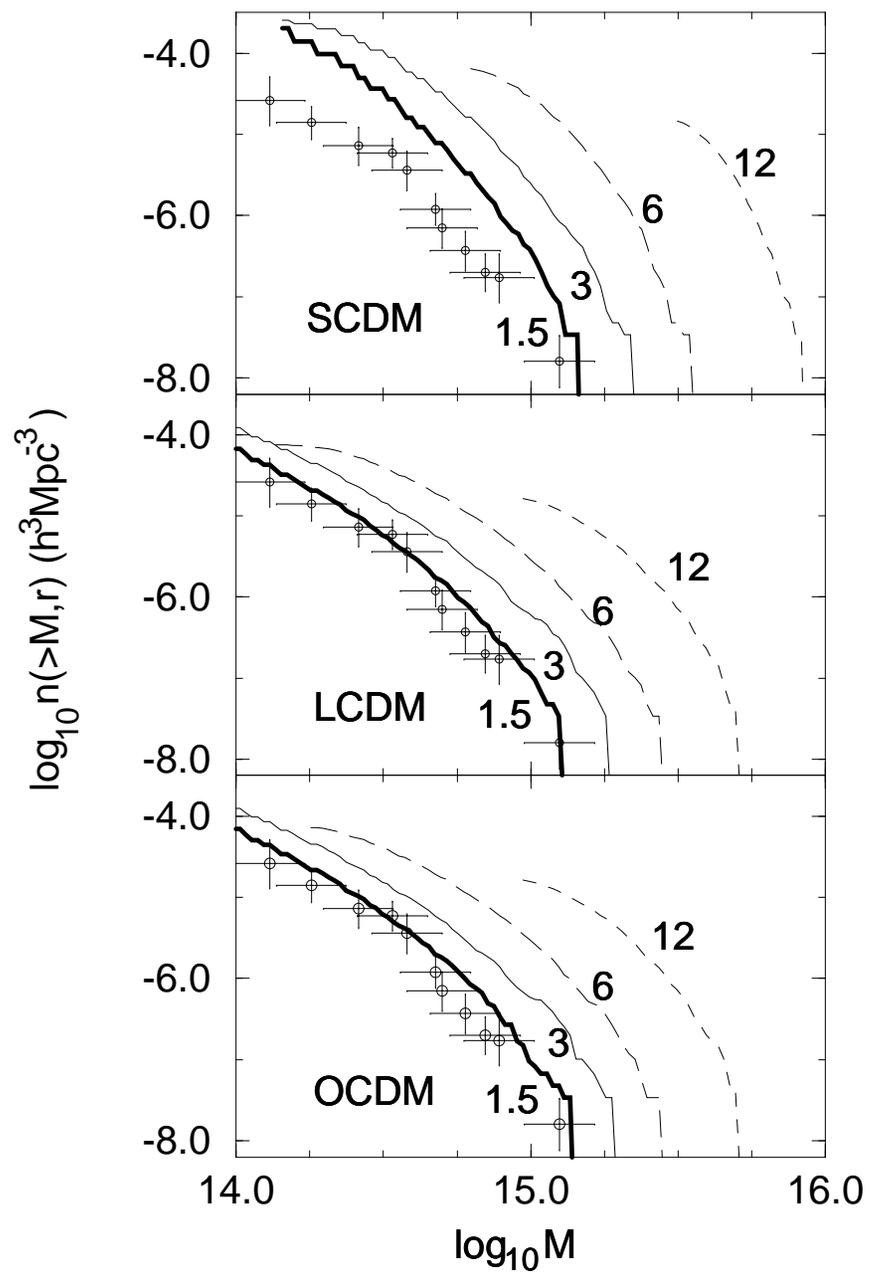
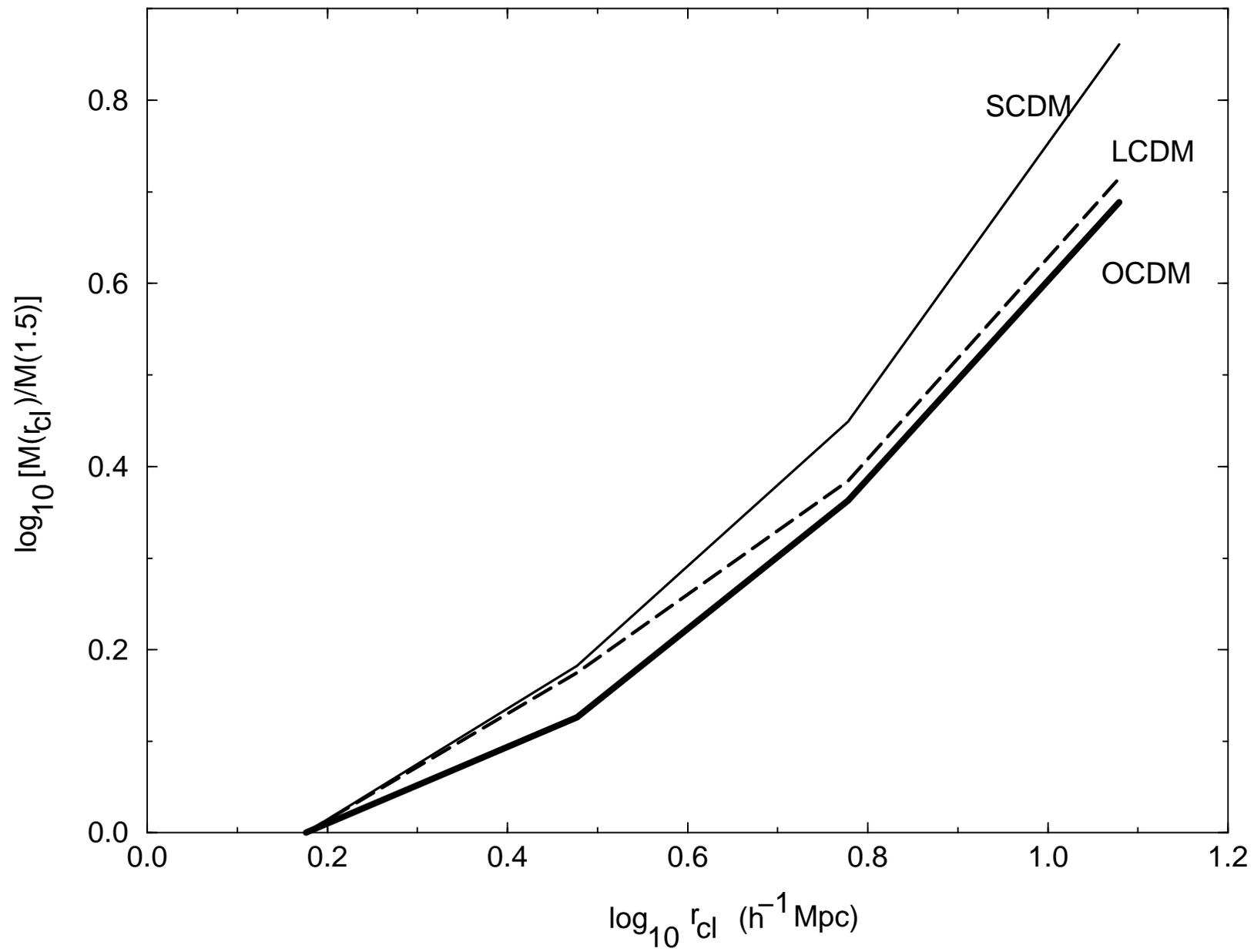
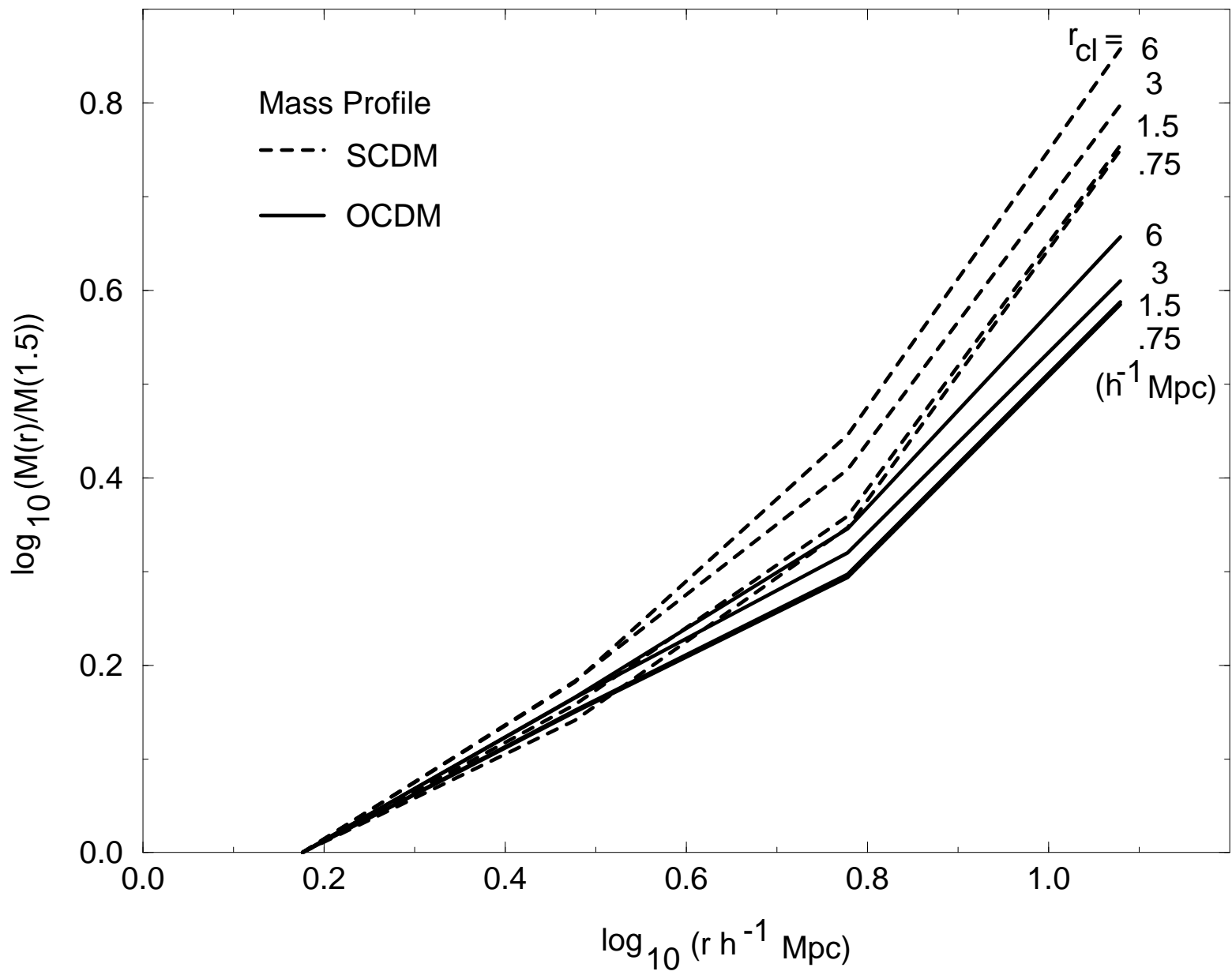
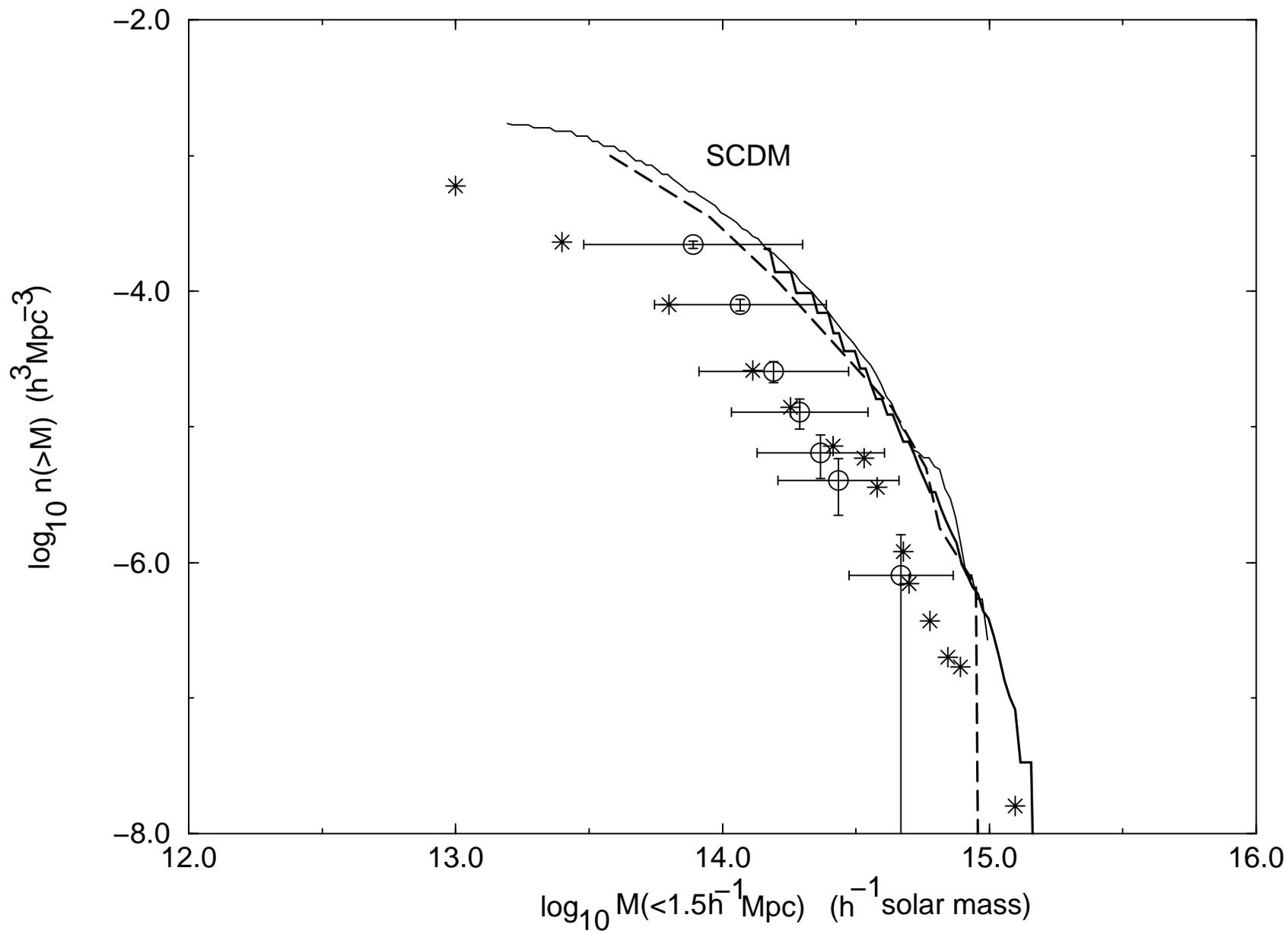
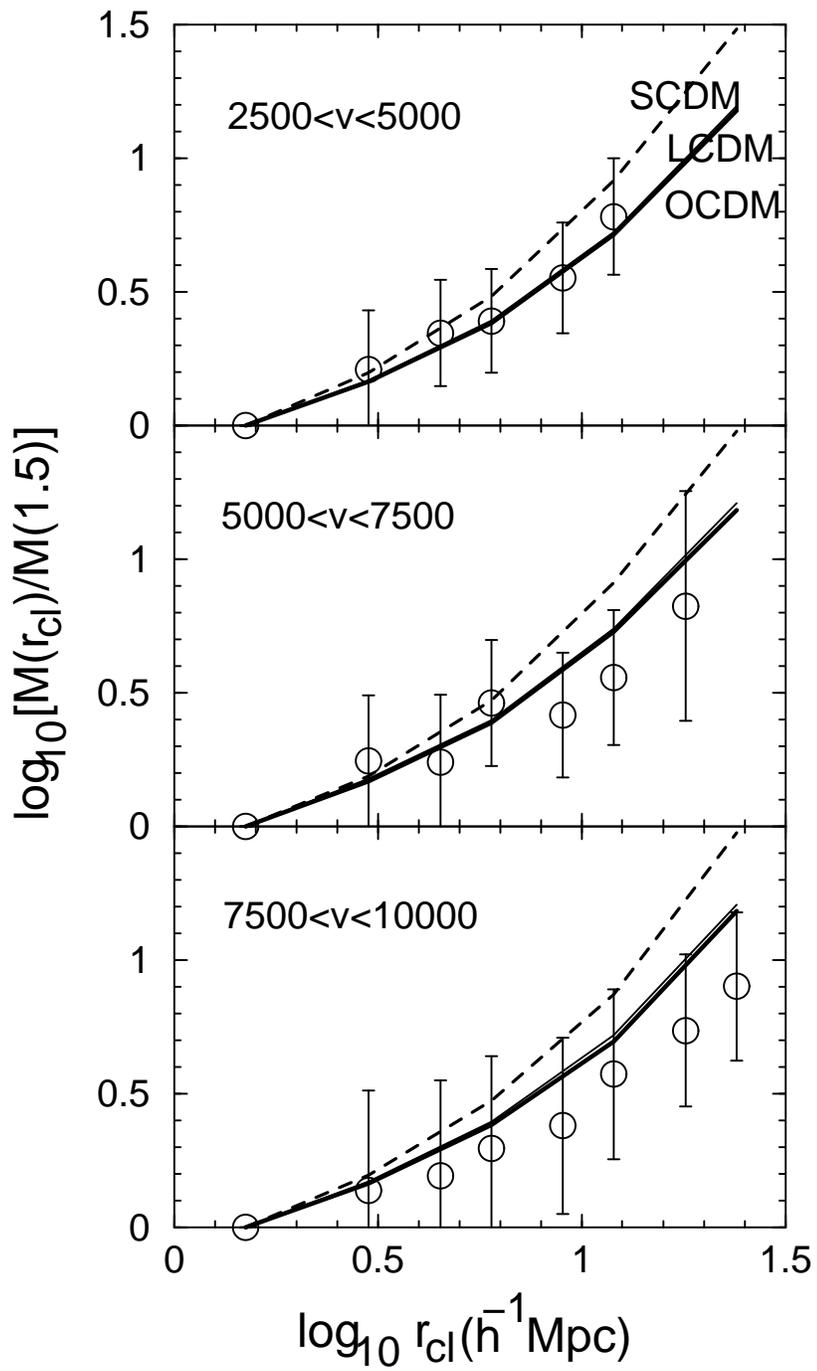
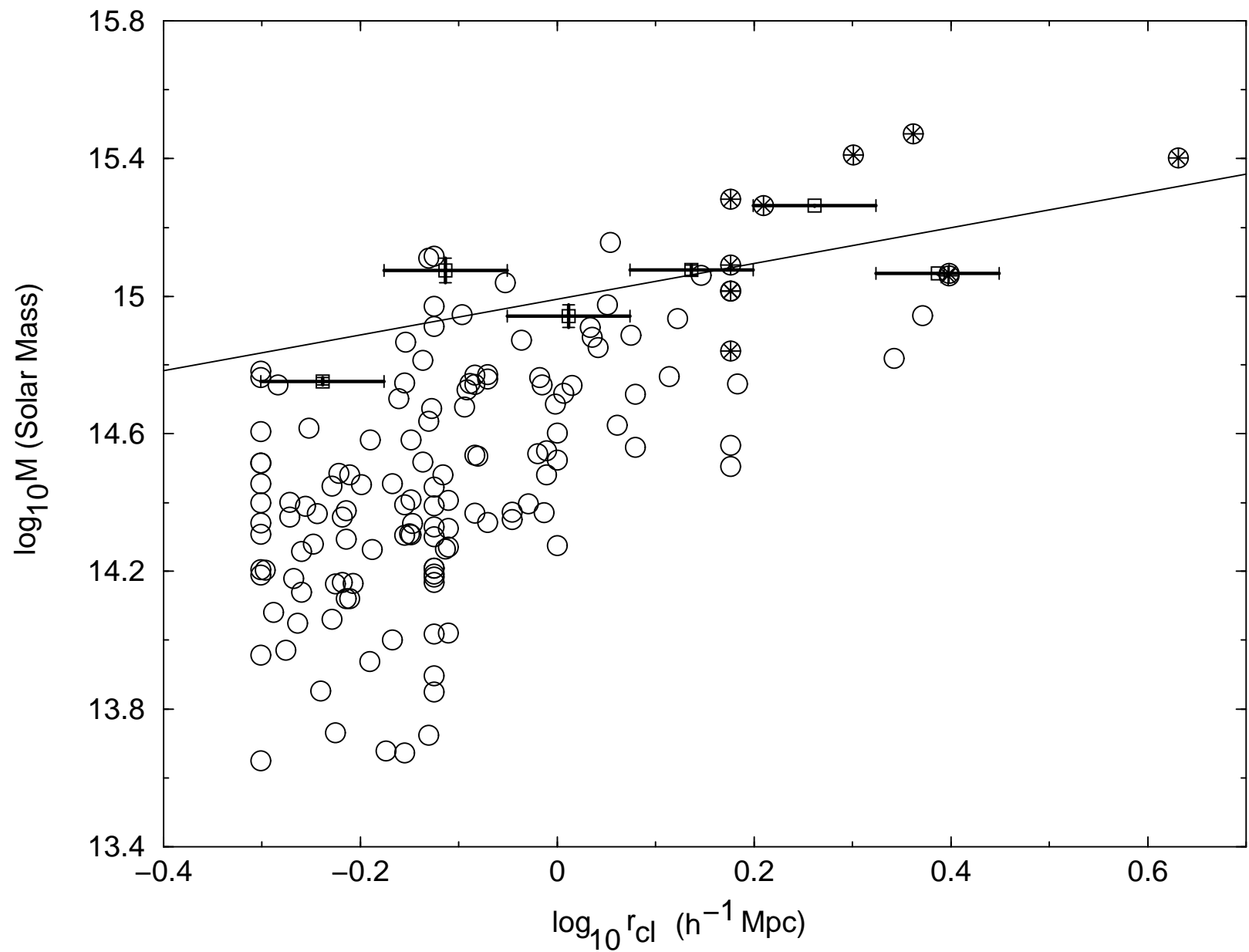
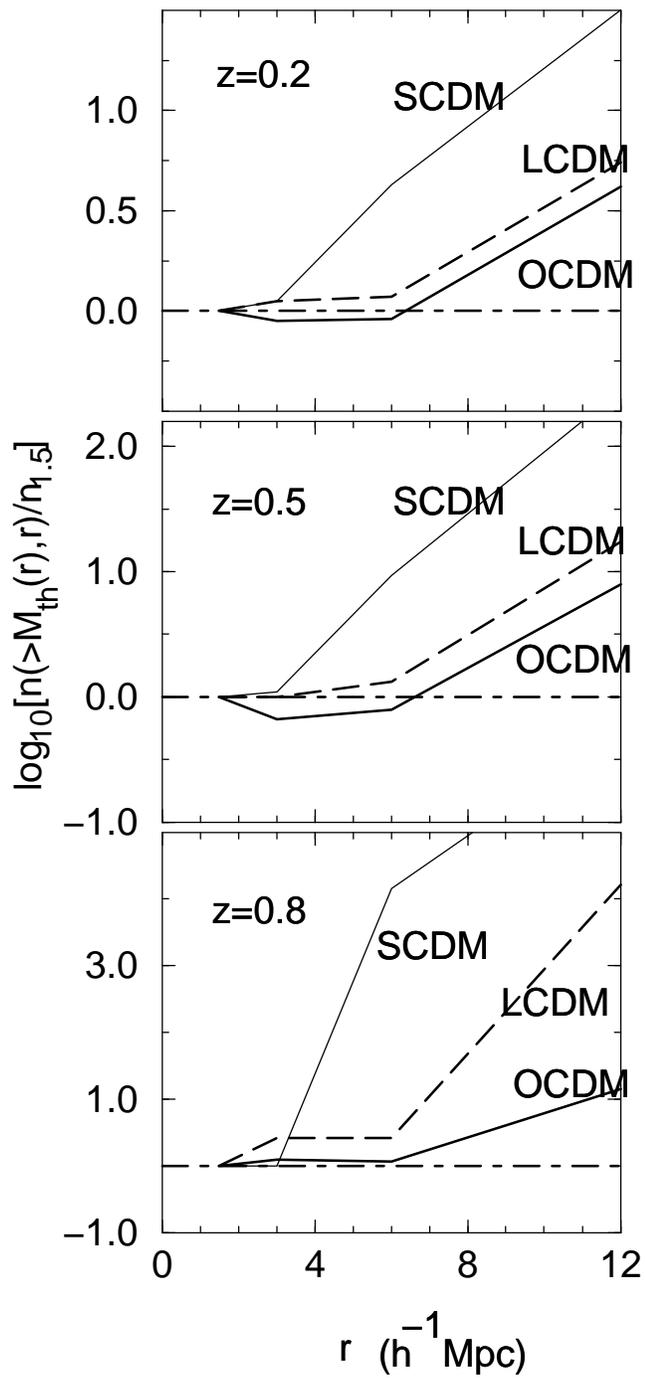
\end{document}